\documentclass[12pt]{article}
\usepackage{amstex,epsfig}
\newcommand{\beq}{\begin{eqnarray}}
\newcommand{\eeq}{\end{eqnarray}}
\begin{document}
\title{Sigma/Glueball Decay of D$^+$ and D$^+_s$}
\author{Leonard S. Kisslinger$^\dagger$\\
        Department of Physics, University of Maryland, College Park, MD 20742}
\maketitle
\indent
\begin{abstract}
   Recently the D$^+$ charm meson was observed to have a clear branching
ratio into the low energy $\pi-\pi$ sigma resonance, while this channel
was not detected in the D$_s^+$ decay. It is shown that this is consistent
with the standard treatment of exclusive charm meson decays and a
proposed glueball/sigma picture.
\end{abstract}

\vspace{0.5 in}

\noindent
PACS Indices:13.25.Ft,14.40.Cs,12.38.Lg,14.40.-n,14.70.Dj,13.75.Lb

\vspace{0.5 in}

  The Fermilab E791 Collaboration has recently reported seeing a clear
signal of the decay of the D$^+$ into a $\pi^+$ plus a low-energy 
scalar-isoscalar $\pi-\pi$ resonance, with a (mass,width) of about 
(480,325) MeV\cite{e791a}. This is presumably the elastic resonance
found\cite{zb} in the analysis of  $\pi-\pi$ scattering with (mass,width) 
of about (400,400) MeV. We refer to this as the sigma, $\sigma$. In the 
D$^+_s$ decays, however, the sigma was not found\cite{e791b}. It is the 
purpose of the present work to show that these experimental results are 
consistent with the glueball/sigma model which we have recently proposed.
With this model there is a unique skeletal diagram which allows us to
calculate the ratio of the $\sigma \pi^+$ to the $\phi \pi^+$ decay widths of 
the D$^+$ with no new parameters. The QCD corrections to the skeletal 
diagrams for these decay channels are then discussed. 
For the D$^+_s$ decay into
the $\sigma \pi+$ channel it is shown that there are two skeletal diagrams
then tend to cancel, which can explain why this channel has not been
observed in experiments.

   The sigma glueball idea is based in part on the QCD sum rule analysis 
of the scalar gluon correlator and in part from the large $\sigma$ decay 
rates of scalar glueball candidates. 
Although the QCD sum rule method is not used in the present work, we 
briefly review the method for finding glueball masses.
Using the composite field operator for the scalar glueball,
$J^G(x) = \alpha_s G^2$, where $G_{\mu\nu}^a$ is the gluon field tensor and
$J^G(x)$ is proportional to the pure-glue term in the QCD Lagrangian, the
scalar gluon correlator is defined as
\beq
\label{1}
\Pi (p) & = & i\int d^4 x \; e^{iq \cdot x}<0 \mid T[J(x) J(0)] \mid 0> \; ,
\eeq
where the link operator has been omitted. The correlator is evaluated
numerically in lattice gauge calculations. In the QCD sum rule method
the dispersion relation for the correlator is equated to a QCD operator
product expansion, with the mass of the Glueball determined by estimating
the value of the pole in the dispersion relation.
Using a subtracted dispersion relation and carrying out the standard QCD
sum rule analysis\cite{nsvz} a number of theorists find a glueball
solution in the region of 300-600 MeV\cite{gb},
depending on the values used for the nonperturbative condensate parameters.
In a recent study of the coupled scalar mesons and glueballs in which 
instanton as well as other gluonic effects have been included\cite{kj}
we find a mainly meson solution corresponding to the f$_o$(1370), a mainly 
glueball solution corresponding to the  f$_o$(1500),  and a light glueball 
in the region of 400-500 MeV\cite{kj}, consistent with our earlier studies.
The main two sources of error in the method are the determination of
the values of the condensates and the treatment of the continuum part of
the dispersion relation. A light glueball has not been found in quenched
lattice gauge calculations.

  The other observations which lead to the glueball/sigma picture are
first that the f$_o$(1500), which has characteristics of a glueball with
a scalar meson admixture, has the four-pion channel as its largest decay 
branching fraction\cite{cbc}; and second, that a BES analysis\cite{yz} has 
shown that the four-$\pi$ channel is dominated by two-sigmas. This is also
true for the f$_o$(1710) and f$_o$(2100) which have recently been shown to
have glueball characteristics. This has led 
to our conjecture that the light glueball and sigma form a coupled-channel
system, with the sigma resonance driven by the glueball pole. From the
Breit-Wigner resonance fit to the resonance\cite{zb} we obtain the matrix
element $<\sigma \mid V^{int} \mid GB>$ for the coupling of the scalar
glueball to the scalar $\pi-\pi$ resonance. Note that the coupling 
interaction $ V^{int}$ is not needed in this work, only the matrix element
taken from the experimental analysis. We have used this for the study
of hybrid baryon decay\cite{lsk1}, and the pomeron-nucleon coupling
and the production of sigmas in high energy p-p collisions via the 
pomeron\cite{lsk2}. This is the glueball/sigma model. Since it is possible
that the treatment of the gluonic continuum could give a false solution
for the light glueball, we consider this picture to be a conjecture,
however, the model would also follow from a very broad gluonic structure
in the region of the $\sigma$ (i.e., the glueball pole is far from the
real axis) which couples strongly to the sigma resonance. If the picture
is valid there are many important consequences that can be
observed in a variety of experiments. We believe that the charm meson
decays are an excellent example, which is the subject of the present
work.

  It is convenient use the quark-diagram classification scheme\cite{cc},
in which there are six skeltal diagrams, by which we mean processes without
explicit gluonic effects. The need to consider processes in addition to what
was once considered to be the dominant spectator decay process was made 
evident by the large difference between D$^+$ and D$^0$ lifetimes\cite{ann}.
See Ref \cite{cc1} for a review of exclusive D$^+$, D$^o$
and D$^+_s$ charm decays expressed in terms of this classification. 
Theoretical treatments have been mainly based on an effective weak Hamiltonian
based on the standard model (see Refs.\cite{bgr,bsw} for reviews of the
method). For example, for the decays of the D$^+$ to a
$\phi + \pi^+$ or $\sigma + \pi^+$, which are of central interest for the
present work, the effective Hamiltonian is
\beq
\label{2}
    H^{eff} & = & \frac{G_F}{\sqrt{2}}V^{*}_{cs}V_{q_e q_f}[(C_1 (\bar{q}_e
 q_f)_L (\bar{s}c)_L +C_2(\bar{s}q_f)_L (\bar{q}_e c)_L],
\eeq
where $(\bar{q}_e q_f)_L = \bar{q}_e^a J_\mu q_f^a \equiv 
\bar{q}_e^a \gamma^\mu (1-\gamma_5) q_f^a$,
$V_{ij}$ are the CKM matrix elements and the color is summed. 
The constants $C_1,C_2$ have been
estimated by renormalization group calculations\cite{rg}. Calculations
with this framework proved to be quite successful for the study of most
exclusive D decays. During the period when the method was being developed,
and many exclusive charm decays were being measured, however, it was
argued that additional flavor singlet gluonic processes, called hairpins,
might be important\cite{cc,hair}. It was also shown that final state 
interactions must be considered on the same level as the hairpins\cite{don},
and that strong interaction effects at least for certain exclusive decays
make it difficult if not impossible to identify the contributions of the
varous skeletal quark diagrams. For inclusive processes one can carry out
operator product expansions and treat the nonperturbative QCD effects using
known condensates (see, e.g., Ref \cite{bb} for a recent review), however,
for exclusive processes accurate inclusion of QCD is difficult.
 
 For the decays D$^+\,\rightarrow\,\phi\,\pi^+$ and  
D$^+\,\rightarrow\,\sigma\,\pi^+$ there is a unique skeletal diagram, the
internal W$^+$ mechanism illustrated in Fig. 1. 
\begin{figure}
\begin{center}
\epsfig{file=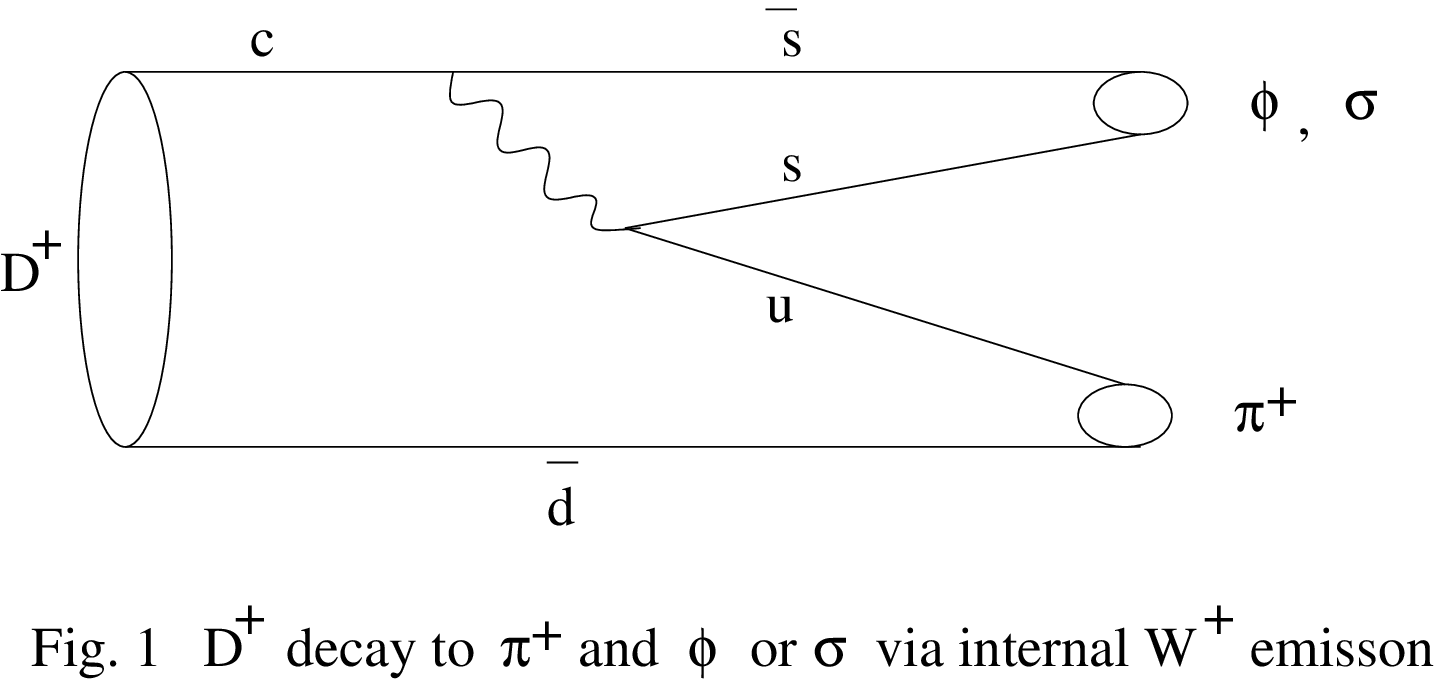,width=12cm}
{\label{Fig.1}}
\end{center}
\end{figure}
Using Eq.(\ref{3}) with a Fierz transformation of the second term one has
\beq
\label{3}
   <X \pi^+ \mid H^{eff} \mid D^+> & = &  \frac{G_F}{\sqrt{2}}
 cos\theta_c sin\theta_c C_3
 <X \mid \bar{s}s)_L \mid 0><\pi^+ \mid (\bar{u} c)_L \mid D^+>.
\eeq
The constant $C_3$ will not be needed here. The matrix elements of the
quarks currents for $X = \phi, \sigma$ are
\beq
\label{4}
 <0 \mid \bar{s}J_\mu s \mid \phi_\nu> & = & f_\phi m_\phi \epsilon_{\mu\nu}\\
\nonumber
  <0 \mid \bar{s}J_\mu s \mid \sigma> & = & f_\sigma m_\sigma \delta_{\mu 0},
\eeq
with $f_\phi, f_\sigma$ the couplings given by the short-distance values of the
wave functions of the $\phi, \sigma$, respectively. For the decay
D$^+\,\rightarrow\,\sigma\,\pi^+$ there is additional contribution from the
process with the $\bar{s}$ and $s$ from the two W$+$ vertices being replaced
by $\bar{d}$ and $d$. The CKM matrix elements are about the same and
$<0\mid\bar{d}J_\mu d\mid\sigma> \simeq 1.2 <0\mid\bar{s}J_\mu s\mid\sigma>$,
since the s quark condensate is about 80\% of the d quark condensate. Thus the
effective value of $f_\sigma$ is increased by about a factor of two.

  Here we use for the value of the $\phi$ coupling\cite{bgr} $f_\phi$ =
228 Mev. The glueball amplitude constant can be estimated from
the low-energy theorem\cite{nsvz}
\beq
\label{5}
\Pi (0) & = & \frac{7}{2} <0 \mid G^2 \mid 0> \; ,
\eeq
with $ <0 \mid G^2 \mid 0>$ the gluon condensate. 
The $\sigma$ coupling can be obtained from another low energy 
theorem\cite{nsvz}
\beq
\label{6}
    i \int d^4x <0 \mid T[\bar{s}s(x)\alpha_s G^2(0)] \mid 0> & \simeq &
  24 \pi <0 \mid \bar{s}s \mid 0>/b_0  \; ,
\eeq
with $b_0$ = 9.667 for three colors and two flavors.
and Eq.(\ref{5}), with a similar form for the d-quark contribution to give
\beq
\label{7}
  <0\mid\bar{s}J_\mu s\mid\sigma>+ <0\mid\bar{d}J_\mu d\mid\sigma> & = & 
  0.14 \delta_{\mu 0} GeV^2.
\eeq
Taking the sigma mass to be 500 Mev, the phase space ratio without spin
$\sigma/\phi$ = 1.23. Therefore from Eq.(\ref{3}) we obtain for the ratio
of the exclusive decay widths
\beq
\label{8}
 \frac{\Gamma(D^+\rightarrow\sigma \pi^+)}{\Gamma(D^+\rightarrow\phi \pi^+)}
   & \simeq & 0.15.
\eeq
The experimental value of this ratio from Ref.\cite{e791a} for the $\sigma$
and the PDG\cite{pdg} for the $\phi$ is 0.22. In the light of both the
uncertainties in the calculation and the possibilitiy of other contributions,
discussed next, there is quite reasonable agreement between theory and
experiment.

   Although the diagram shown in Fig.1 is the only skeletal diagram
that contributes to either the $\phi$ or $\sigma$ decays of the D$^+$,
there are two other diagram with explicit QCD processes that also contribute:
\begin{figure}
\begin{center}
\epsfig{file=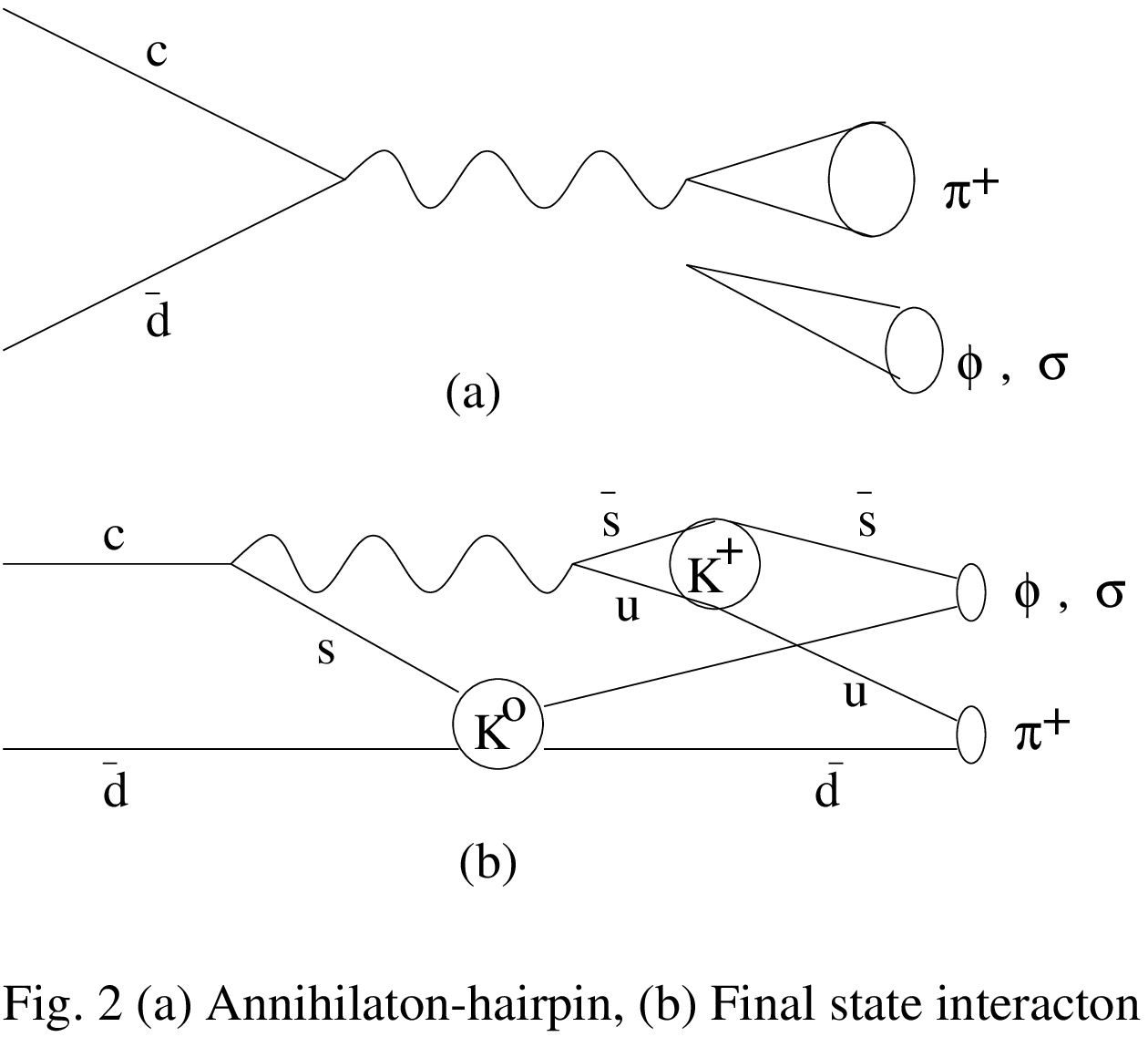,width=12cm}
{\label{Fig.2}}
\end{center}
\end{figure}
the annihilation-hairpin processes (Fig.2a) and the spectator-final state
interaction process (Fig. 2b). The later is analogous to the process 
treated in Ref.\cite{don} in which it was shown that such final state
rearrangement processes might be comparable in magnitude to the nonspectator
decays. We do not attempt to estimate these prosesses which are very 
difficult to calculate accurately.

   For the D$_s^+$ decay into a $\phi \pi^+$ there is one skeletal diagram,
the spectator shown in Fig.3a; while for the decay into a $\sigma \pi^+$
there is a contribution from the spectator graph and also two annihilation
graph processes, shown in Fig.3b. 
\begin{figure}
\begin{center}
\epsfig{file=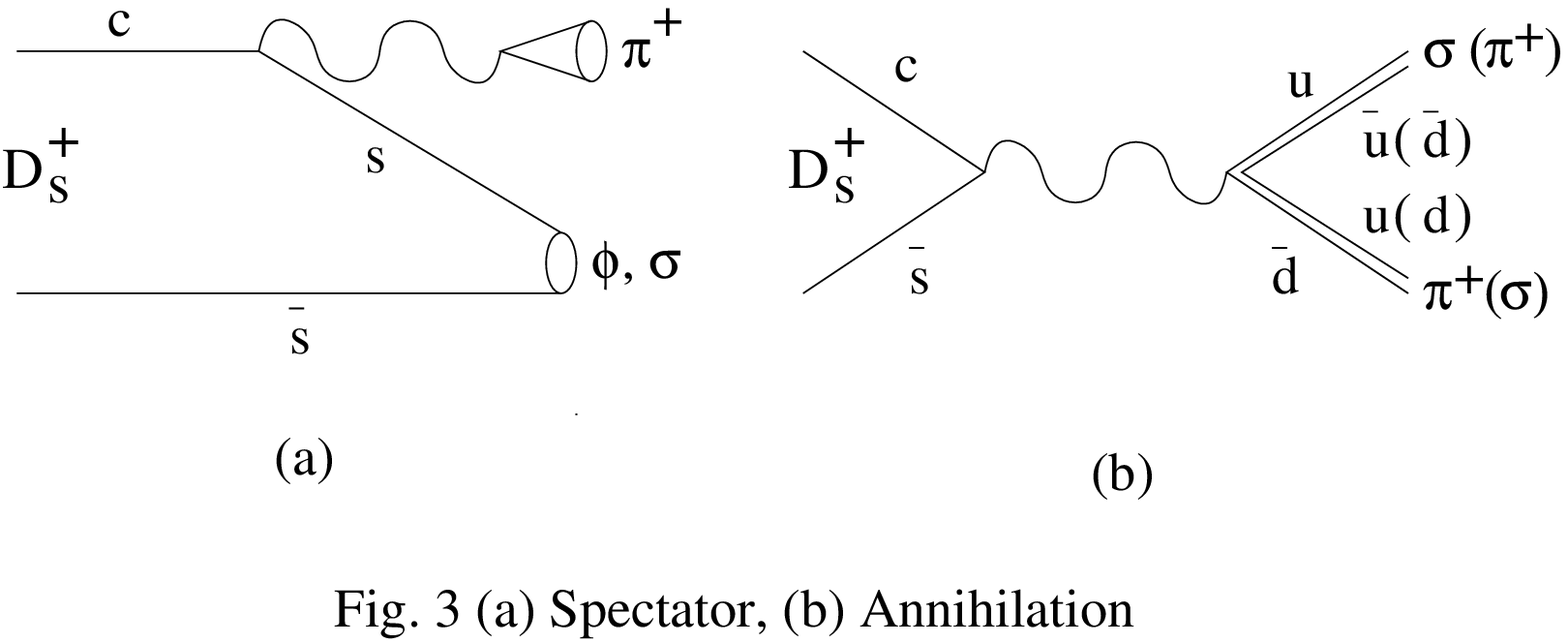,width=12cm}
{\label{Fig.3}}
\end{center}
\end{figure}
For the $D_s^+ \rightarrow \sigma \pi^+$ the matrix element of the weak 
effective Hamiltonian, after a Fierz transformation, is
\beq
\label{9}
   <\sigma \pi^+ \mid H^{eff} \mid D_s^+> & = &  \frac{G_F}{\sqrt{2}}
 cos\theta_c^2 (C_1+3 C_2)
 <\sigma \pi^+ \mid (\bar{u}d)_L \mid 0><0 \mid (\bar{s} c)_L \mid D_s^+>.
\eeq
Using the results given in Ref.\cite{bgr} one sees that the ratio of the $\sigma$
to $\phi$ rates for the D$_s^+$ decays, which contains a factor of 
$(C_1 + 3 C_2)^2/C_1^2$, would be strongly reduced compared to the D$^+$ decay.
This gives a qualitative explanation for the fact that the $\sigma \pi^+$
channel was not found\cite{e791b} in the E791 experiment.

  It is interesting to note that the only skeletal diagram for the
D$^+ \rightarrow K^+ + \phi (\sigma)$ decays is the annihilation process;
and that there are also two D$^o$ decays to the $\sigma$ or $\phi$,
D$^o \rightarrow \phi(\sigma) \bar{K}^o$ and D$^o \rightarrow \phi(\sigma) 
K^o$, that
have the W-exchange as the only skeletal diagram. We expect that the ratio 
of the $\sigma$ to the $\phi$ fraction
should be similar to the corresponding D$^+$ decays described above. 
Another interesting related experimental observation by the CLEO 
Collaboration\cite{cleo} is that in the
$\tau^- \rightarrow \nu_\tau \pi^- \pi^o \pi^o$ decay the data cannot be
fit without a sigma channel.  The diagram with the W$^-$ coupling to the
\begin{figure}
\begin{center}
\epsfig{file=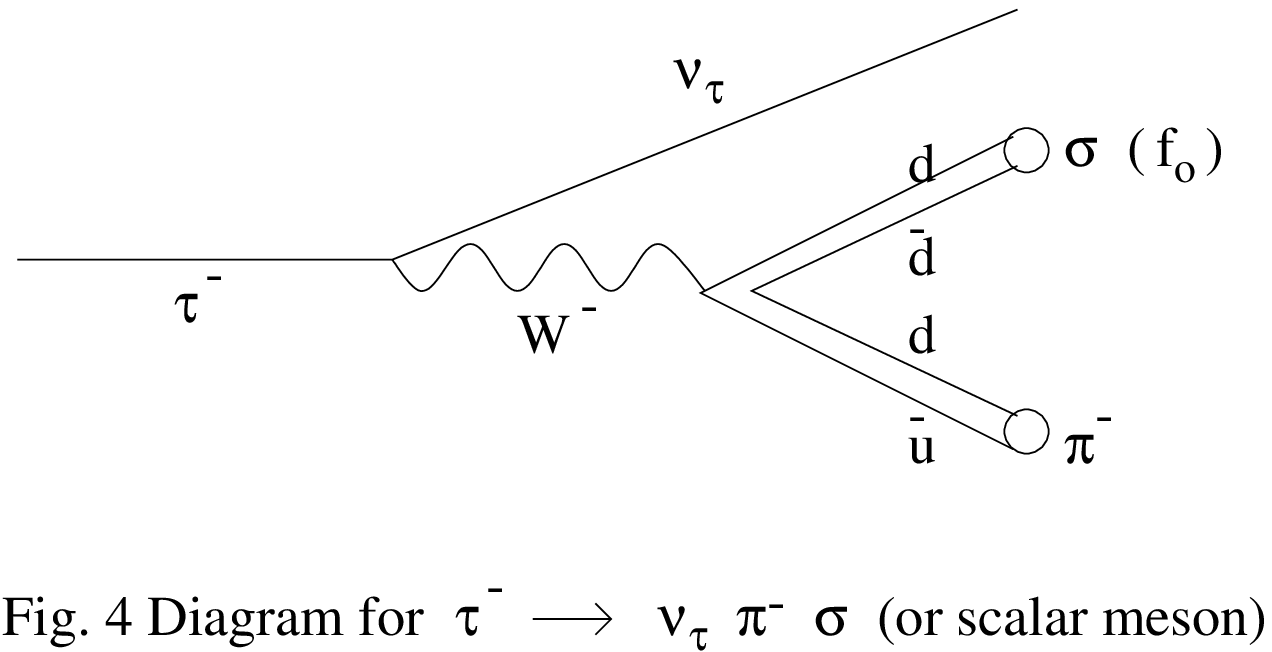,width=12cm}
{\label{Fig.4}}
\end{center}
\end{figure}
$ \sigma \pi^-$ through a $d \bar{u}$, illustrated in Fig.4,
leads to this channel in a natural way in the glueball/sigma model. In this 
case a more detailed theoretical study of the $(\bar{q}_a q_b)_L$ matrix
elements is needed to obtain the branching ratios, which we do not attempt 
here.

We conclude that the D$^+$ and  D$_s$ decays are consistent with our 
glueball/sigma model and suggest that experimental searches for the
sigma channels in other charm decays, as well as other heavy quark and
lepton hadronic decays, would be rewarding.

$^\dagger$ On leave from Carnegie Mellon University

 This work was supported in part by the NSF grant PHY-00070888. The author
wishes to thank the TQHN group at the University of Maryland for hospitality 
during the period when this work was carried out, and acknowledge helpful
discussions with Jen-chieh Peng, Roy Briere, Jeffrey Appel and Thomas Cohen.

\end{document}